\documentclass{elsarticle}
\usepackage[utf8]{inputenc}
\usepackage[margin=1.5cm]{geometry}
\usepackage{titlesec}
\usepackage{enumitem}
\usepackage{amssymb}
\usepackage{xcolor}
\usepackage{acronym}
\usepackage{soul}
\usepackage{booktabs}
\usepackage[normalem]{ulem}
\newlist{selectlist}{itemize}{2}
\setlist[selectlist]{label=$\square$,leftmargin=*,noitemsep,topsep=0pt}

\usepackage{lmodern}

\usepackage{hyperref}
\hypersetup{
    colorlinks=true,
    linkcolor=blue,
    filecolor=magenta,      
    urlcolor=blue,
}
 
\titleformat{\section}[block]{\hspace{1em}\bfseries}{\thesection.}{0.5em}{} 
\titleformat{\subsection}[block]{\hspace{1em}}{\thesubsection}{0.5em}{}

\usepackage[dvipsnames]{xcolor}
\usepackage{listings}
\usepackage{bm}
\usepackage{adjustbox}

\definecolor{DeepGreen}{rgb}{0.0,0.5,0.0}

\usepackage{soul}
\usepackage{multicol}
\usepackage{multirow}
\usepackage{orcidlink}

\definecolor{bgcolor}{rgb}{0.95,0.95,0.92}  % Light gray background

\journal{SoftwareX}

\begin{document}

% \begin{acronyms}
    \newacro{ai}[AI]{Artificial Intelligence}
    \newacro{a2c}[A2C]{Advantage Actor-Critic}
    \newacro{cdf}[CDF]{Cumulative Distribution Function}
    \newacro{cagr}[CAGR]{Compound Annual Growth Rate}
    \newacro{psm}[PSM]{Power State Management}
    \newacro{psmp}[PSMP]{\ac{psm} Policies}
    \newacro{cl}[CL]{Curriculum Learning}
    \newacro{drl}[DRL]{Deep \ac{rl}}
    \newacro{dnn}[DNN]{Deep Neural Networks}
    \newacro{fcfs}[FCFS]{First Come First Served}
    \newacro{hpc}[HPC]{High Performance Computing}
    \newacro{mdp}[MDP]{Markov Decision Process}
    \newacro{mq}[MQ]{Message Queue}
    \newacro{dtmdp}[DTMDP]{Discrete-Time \ac{mdp}}
    \newacro{ctmdp}[CTMDP]{Continuous-Time \ac{mdp}}
    \newacro{ml}[ML]{Machine Learning}
    \newacro{pdf}[PDF]{Probability Density Dunction}
    \newacro{psm}[PSM]{Power State Management}
    \newacro{psmp}[\ac{psm}P]{\ac{psm} problem}
    \newacro{qos}[QoS]{Quality of Service}
    \newacro{rmse}[RMSE]{Root Mean Squared Error}
    \newacro{rtl}[RTL]{Real-Time Learning}
    \newacro{rl}[RL]{Reinforcement Learning}
    \newacro{swf}[SWF]{Standard Workload Format}
    \newacro{api}[API]{Application Programming Interface}
    \newacro{ipc}[IPC]{Inter-Process Communication}
    \newacro{zmtp}[ZMTP]{ZeroMQ Message Transport Protocol}
    \newacro{psas}[PSAS]{Power State-Aware Scheduler}
    \newacro{ipm}[IPM]{Intelligent Power Manager}
    \newacro{psasipm}[PSAS+IPM]{\ac{psas} + \ac{ipm}}
    \newacro{psasao}[PSAS (Auto On)]{\ac{psas} with Auto Switch On}
    \newacro{psus}[PSUS]{Power State-Unaware Scheduler}

\begin{frontmatter}
% \dochead{}

\title{SPARS: A Reinforcement Learning-Enabled Simulator for Power Management in HPC Job Scheduling}

\author[1]{Muhammad Alfian Amrizal\,\orcidlink{0000-0003-1124-5137}}\ead{muhammad.alfian.amrizal@ugm.ac.id}
\author[2]{Raka Satya Prasasta\,\orcidlink{0009-0006-5751-6050}}\ead{2200018273@webmail.uad.ac.id}
\author[1]{Santana Yuda Pradata\,\orcidlink{0009-0009-0470-8456}}\ead{santanayudapradata@mail.ugm.ac.id}
\author[3,4]{Kadek Gemilang Santiyuda\,\orcidlink{0000-0002-4432-6059}}\ead{gemilang.santiyuda@instiki.ac.id}
\author[1]{Reza Pulungan\corref{cor1}\,\orcidlink{0000-0002-5019-1357}}\ead{pulungan@ugm.ac.id}
\author[5]{Hiroyuki Takizawa\,\orcidlink{0000-0003-2858-3140}}\ead{takizawa@tohoku.ac.jp}

\address[1]{Department of Computer Science and Electronics, Universitas Gadjah Mada, Yogyakarta 55281, Indonesia}
\address[2]{Faculty of Industrial Technology, Universitas Ahmad Dahlan, Yogyakarta 55191, Indonesia}
\address[3]{Pascasarjana, Institut Bisnis dan Teknologi Indonesia, Bali 80225, Indonesia}
\address[4]{Department of Industrial Management, National Taiwan University of
Science and Technology, Taiwan 106335, Taiwan}
\address[5]{Cyberscience Center, Tohoku University, Sendai 9800845, Japan}
\cortext[cor1]{Corresponding author.} 

\begin{abstract}
High-performance computing (HPC) systems consume enormous amounts of energy, with idle nodes as a major source of energy waste. Powering down idle nodes can mitigate this problem, but long boot/shutdown delays can introduce significant queueing penalties if transitions are poorly timed. To address this trade-off, we present SPARS, a reinforcement learning-enabled simulator for power management in HPC job scheduling. SPARS integrates job scheduling and node power-state management within a discrete-event simulation framework. It supports traditional scheduling policies such as First Come First Serve and EASY Backfilling, along with enhanced variants that employ reinforcement learning agents to dynamically decide when nodes should be powered on or off. Users can configure workloads and platforms in JSON format, specifying job arrivals, execution times, node power models, and transition delays. The simulator records comprehensive metrics---including energy usage, wasted power, job waiting times, and node utilization---and provides Gantt chart visualizations to analyze scheduling dynamics and power transitions. Unlike widely used Batsim-based frameworks that rely on heavy inter-process communication, SPARS provides lightweight event handling and consistent simulation results, making experiments easier to reproduce and extend. Its modular design allows new scheduling heuristics or learning algorithms to be integrated with minimal effort. By providing a flexible, reproducible, and extensible platform, SPARS enables researchers and practitioners to systematically evaluate power-aware scheduling strategies, explore the trade-offs between energy efficiency and performance, and accelerate the development of sustainable HPC operations.
\end{abstract}


\end{frontmatter}

\section*{Code metadata}
\vspace{-20pt}
% \noindent
% \textbf{Code metadata}\\
% \textit{Please replace the italicized text in the right column with the correct information about your code/software and leave the left column untouched.}\\

\begin{table}[!htp]
  \centering
  \caption{Code metadata.}
  \label{codeMetadata}   
  \begin{adjustbox}{max width=\textwidth}
    \begin{tabular}{lp{8cm}p{8.5cm}}
      \toprule
      No. & Code metadata description & Information \\
      \midrule
C1 & Current code version & v1\\
C2 & Permanent link to code/repository used for this code version & \url{https://github.com/RakaSP/SPARS-Pub}\\
C3 & Legal code license & MIT license\\
C4 & Code versioning system used & git\\
C5 & Software code languages, tools, and services used & Python\\
C6 & Compilation requirements, operating environments \& dependencies & Linux OS, Windows OS, Python 3.11.13, gymnasium 1.2.0, torch 2.5.1+cu121, torchvision 0.23.0+cpu, CUDA drivers in OS, etc.\\
C7 & If available link to developer documentation/manual & \url{https://github.com/RakaSP/SPARS-Pub/blob/main/README.md}\\
C8 & Support email for questions & \href{mailto:pulungan@ugm.ac.id}{pulungan@ugm.\allowbreak{}ac.\allowbreak{}id}\\
      \bottomrule
    \end{tabular}
  \end{adjustbox}
\end{table}

% \textcolor{red}{
% \textit{Optionally, you can provide information about the current executable
% software version filling in the left column of
% Table~\ref{executabelMetadata}. Please leave the first column as it is.}
% }

% \begin{table}[!h]
% \begin{tabular}{|l|p{6.5cm}|p{6.5cm}|}
% \hline
% \textbf{Nr.} & \textbf{(Executable) software metadata description} & \textbf{Please fill in this column} \\
% \hline
% S1 & Current software version & \textcolor{red}{For example 1.1, 2.4 etc.} \\
% \hline
% S2 & Permanent link to executables of this version  & \textcolor{red}{For example: \url{https://github.com/combogenomics/DuctApe/releases/tag/DuctApe-0.16.4}} \\
% \hline
% S3  & Permanent link to Reproducible Capsule & \\
% \hline
% S4 & Legal Software License & \textcolor{red}{List one of the approved licenses} \\
% \hline
% S5 & Computing platforms/Operating Systems & Windows, Linux \\
% \hline
% S6 & Installation requirements \& dependencies & \\
% \hline
% S7 & If available, link to user manual - if formally published include a reference to the publication in the reference list & \textcolor{red}{For example: \url{http://mozart.github.io/documentation/}} \\
% \hline
% S8 & Support email for questions & \href{mailto:flanetern@gmail.com}{flanetern@gmail.com} \\
% \hline
% \end{tabular}
% \caption{Software metadata (optional)}
% \label{executabelMetadata} 
% \end{table}

\section{Motivation and significance}\label{isect1}

High-performance computing (HPC) systems are responsible for an estimated 140 billion kWh of power usage annually~\cite{nrdc}, with operational costs doubling roughly every five years \cite{buyya2013mastering}. A significant portion of this energy is wasted on idle nodes---nodes that are powered on but not executing jobs. Energy-aware HPC scheduling surveys confirm that dynamic power management of idle nodes is one of the most studied and impactful mechanisms for reducing this waste \cite{kocot2023energy}. In practice, a Slurm-based idle-node shutdown policy \cite{yoo2003slurm} at J\"ulich Supercomputing Centre was associated with a 25\% reduction in annual power consumption \cite{suarez2025energy}.

However, powering down idle nodes introduces a fundamental operational challenge. Across various HPC systems, the process of shutting down and powering on nodes often takes a very long time, typically 10--15 minutes~\cite{Ohmura}, and in extreme cases can reach 30--45 minutes~\cite{budiarjo2025improving}. Naively powering off idle nodes can therefore cause newly submitted jobs to wait unnecessarily long for nodes to become available, offsetting the energy savings with significant performance penalties. Designing an effective energy-aware strategy thus requires carefully balancing these two competing objectives---minimizing idle energy waste while avoiding excessive job waiting times.

\begin{figure}[!ht]
  \centering
  \includegraphics[width=0.9\linewidth]{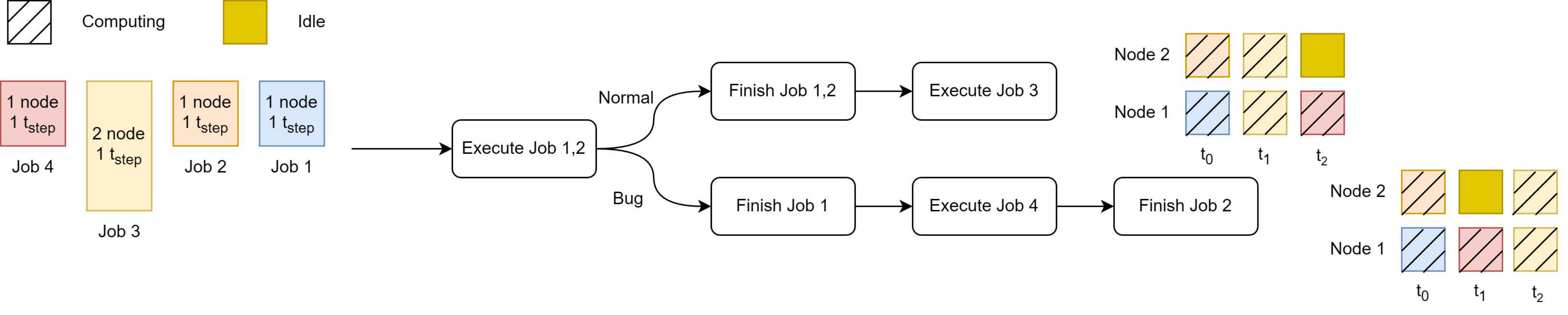}
  \caption{Same-time event batching issue. The example contains four jobs on a two-node system. Jobs~1 and~2~are running on Nodes~1 and~2 and are scheduled to complete simultaneously at $t_1$, while Jobs~3 and~4 wait in the queue. The bug splits these simultaneous completions, causing the scheduler to make different decisions. Note that labels $t_0$, $t_1$, and $t_2$ are event timestamps, not uniform intervals.}
  \label{fig:Bug1}
\end{figure}

Exploring this trade-off directly on production systems is impractical due to scale, cost, and the risk of disrupting ongoing workloads. Therefore, a dedicated simulator is essential to enable controlled experimentation with diverse workloads, machine configurations, and power-transition policies. Through simulation, researchers can evaluate the consequences of different scheduling and power management strategies, quantify their impact on both energy efficiency and job performance, and iteratively refine approaches before deployment on actual HPC systems.

To enable such controlled experimentation, researchers commonly rely on simulation frameworks to study and evaluate new policies before deployment. In the context of HPC batch scheduling, widely used simulators include AccaSim~\cite{galleguillos2020accasim}, which targets extensible HPC batch scheduling, and ElastiSim~\cite{ozden2022elastisim}, which focuses on malleable workload scheduling. However, neither provides native node power-state transitions for studying detailed boot/shutdown dynamics. While such features could, in principle, be added, doing so would require intrusive modifications.

Another widely adopted simulator is Batsim~\cite{dutot2015batsim}. It includes built-in power-state support and provides multiple scheduler interfaces in different languages (e.g., batsched in C$++$ and pybatsim in Python), making it convenient for developing and evaluating a broad range of scheduling policies. However, our study reveals several limitations that can affect the reliability and efficiency of the simulations. First, it relies on inter-process communication, which adds overhead. Second, events occurring at the same simulation timestamp may be delivered in separate scheduler messages rather than as a single atomic batch, which can lead to divergent decisions across logically equivalent runs~\cite{batsim-issue67}.

\hyperref[fig:Bug1]{Fig.~\ref{fig:Bug1}} illustrates how same-time event batching can alter scheduling decisions. At $t_1$, Jobs~1 and~2 complete simultaneously, while Jobs~3 and~4~are in waiting queue. In the normal case, both completions are processed together, and Job~3 starts immediately on two nodes. In the buggy case, the completions are split, so Job~4 is backfilled first, and Job~3 is delayed by one step. This leads to different schedules, utilization, and energy traces for logically equivalent simulations, complicating evaluation of schedulers and {Power-State Management (PSM)} policies, i.e., policies that decide when compute nodes should be switched on, kept active, or powered off.

Beyond the aforementioned Batsim-specific concerns, there is a broader gap in the simulation ecosystem: as power-aware scheduling research increasingly explores learning-based control, many simulators lack a native Reinforcement Learning (RL)/Deep Reinforcement Learning (DRL) \cite{suttonbarto} interface to support closed-loop training.
A systematic review of RL/DRL for data-center energy efficiency highlights that learning-based approaches are often able to optimize operational decisions, such as task scheduling \cite{8906698}, resource allocation \cite{9239288}, virtual machine (VM) consolidation \cite{9698981}, and even facility-side control \cite{kahil2025reinforcement} (e.g., cooling \cite{8772127}). 
Consistent with this trend, RL-enabled simulation setups such as DRAS-CQSim~\cite{fan2021dras} and RLScheduler~\cite{zhang2020rlscheduler} primarily focus on job scheduling decisions. However, they share the same limitation: they do not model the detailed node power-state transitions needed to study idle-node power management.

Despite the growing need for power-aware scheduling in HPC, no existing simulator natively integrates both RL and comprehensive PSM within a unified framework. To fill this gap, this paper proposes SPARS (Simulator for Power-Aware RL Scheduling). SPARS's core contribution is its seamless co-design of a power-state manager and a job scheduler, enabling RL agents to directly control node power transitions while the scheduler manages job dispatch---a capability no prior simulator supports out of the box. SPARS formulates this power-aware scheduling problem as a Markov Decision Process (MDP)~\cite{suttonbarto}, and includes a GYM-compatible \cite{1606.01540} interface that enables training of AI-based models with learning updates triggered after each simulation event, built around scheduling logic inspired by the EASY Backfilling algorithm~\cite{carastan2019one,feitelson1998utilization}.

\section{Software description}\label{Section~2}

SPARS is a Python-based discrete-event simulator designed to model, analyze, and optimize job scheduling dynamics within HPC systems in continuous time. Its primary purpose is to enable researchers and system designers to evaluate scheduling algorithms under configurable workloads and platform settings without relying on real-world HPC deployments.

The simulator operates through an event-driven workflow in which job arrivals, executions, completions, and node state transitions are processed chronologically. Users define workloads and platforms, then select scheduling policies from built-in options, such as First Come First Serve (FCFS) and EASY Backfilling, or their own customized scheduling and PSM policies for evaluation. During simulation, SPARS tracks energy consumption, job waiting times, node utilization, resource waste metrics, and the data needed to visualize scheduling results in a Gantt chart, providing comprehensive insights into the effectiveness of scheduling algorithms.

\subsection{Software architecture}\label{isect2}

\begin{figure}[!ht]
  \centering
  \includegraphics[width=1.0\linewidth]{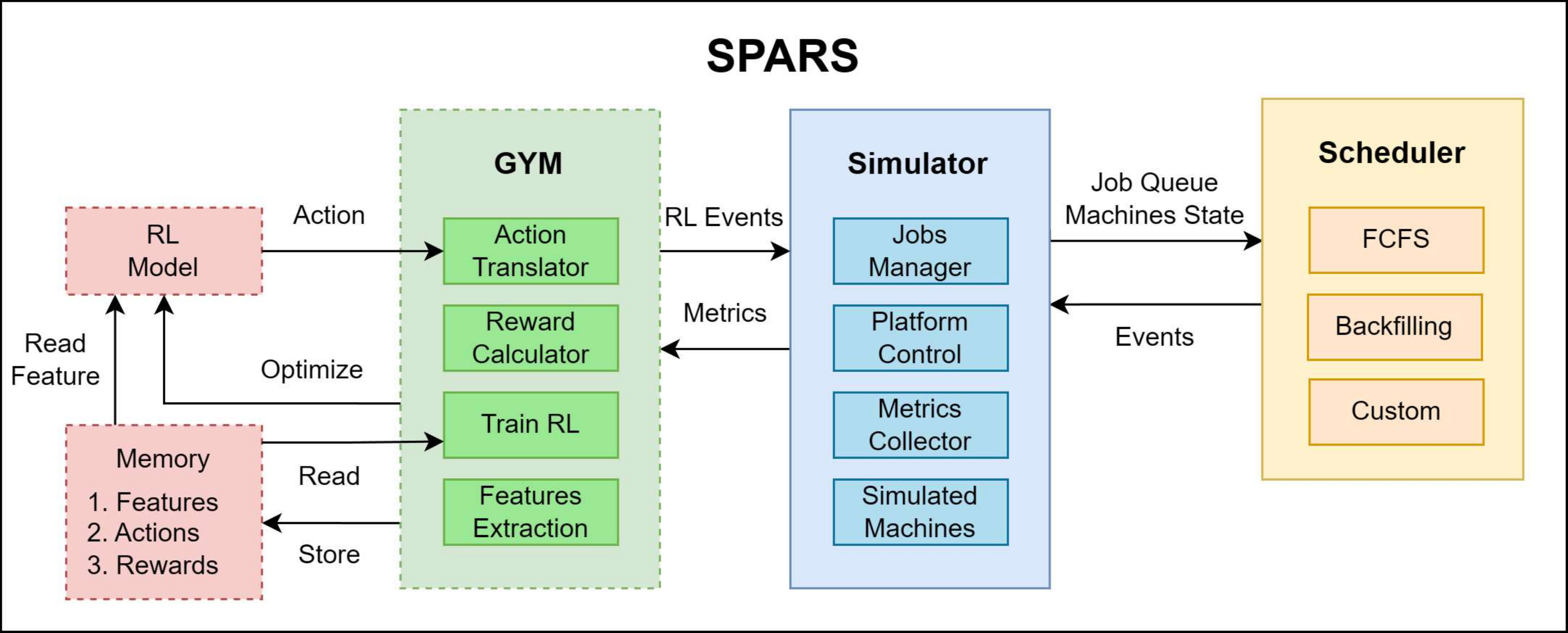}
  \caption{SPARS software architecture. }
  \label{fig:SPARSArchitecture}
\end{figure}

The core architecture of SPARS, depicted in \hyperref[fig:SPARSArchitecture]{Fig.~\ref{fig:SPARSArchitecture}}, consists of an event-driven HPC simulator coupled with an external scheduler. The simulator is organized into four main components: the Jobs Manager, which maintains the job queue and dispatches jobs; the Platform Control, which applies state transitions to the simulated machines (e.g., powering nodes on or off, changing operating states); the Metrics Collector, which records runtime statistics; and the Simulated Machines model. The scheduler observes the current job queue and machine states exported by the simulator and replies with scheduling decisions encoded as events (e.g., job start, cancellation, or node-state changes). SPARS provides several built-in scheduler implementations, including FCFS, EASY Backfilling, and allows users to plug in their own schedulers. The components highlighted with dashed borders in \hyperref[fig:SPARSArchitecture]{Fig.~\ref{fig:SPARSArchitecture}} are optional and implement the RL extension of SPARS.

\subsection{Software functionalities}\label{isect3}

SPARS
unifies workload and platform modeling, scheduling execution, and energy accounting in continuous time. Users define jobs and machines using JSON files that specify attributes such as arrival times, job sizes, durations, node requirements, state transition delays, and power models. Experiments can then be conducted using pluggable schedulers, including FCFS, EASY Backfilling, and other PSM variants from our previous work, such as Power State-Aware Scheduler (PSAS) combined with Intelligent Power Manager (IPM), called PSAS+IPM, PSAS with Auto Switch On (PSAS (Auto On), and Power State-Unaware Scheduler (PSUS)~\cite{workshop}, or RL-based PSM~\cite{budiarjo2025improving,khasyah2022advantage} via a GYM-compatible interface. During a run, the engine processes arrivals, dispatches, completions, and on/off transitions while tracking energy in active, idle, sleep, and transition states. SPARS writes CSV logs for jobs and node states, reports utilization, mean waiting time, total energy, and wasted energy, and renders Gantt charts to visualize scheduling and power behavior. A modular architecture lets researchers swap scheduling logic, feature extractors, reward functions, and learning algorithms without changing the core, which makes policy comparisons reproducible and extensions straightforward.

\subsection{Basic usage of SPARS}\label{isect4}

This section provides a guide to basic SPARS usage. A more comprehensive user guide is available in the simulator's GitHub repository.

\subsubsection{Instantiating a problem case}\label{isect5}

To simulate a scheduling scenario in SPARS, users must define the platform and the workload. Both are described in a JSON file. The \nobreak{}platform describes the hardware specifications of the HPC system, including the number of compute nodes, the node state transition time, and the energy consumption in each state. The platform configuration enables simulation of diverse hardware environments. In \texttt{platform.json}, each node includes a unique identifier (\texttt{id}), Dynamic Voltage and Frequency Scaling (DVFS) profiles (\texttt{dvfs\_profiles}) that map operating modes to nominal power and normalized compute speed, a default DVFS mode (\texttt{dvfs\_mode}), and a power-state model (\texttt{states}) describing the supported states, their power consumption and compute speeds, and the allowed transitions and transition times between them.

The workload defines the set of jobs that will be submitted and scheduled. Each job includes information such as its submission time, the number of compute nodes required, the expected runtime, and a unique job identifier. The \texttt{workload.json} file specifies the job stream with the fields summarized in \hyperref[tab:workload-json]{Table~\ref{tab:workload-json}}.

\begin{table}[htbp]
    \centering
    \caption{Structure of \texttt{workload.json}.}
    \label{tab:workload-json}
    \begin{tabular}{p{0.28\textwidth}p{0.62\textwidth}}
        \toprule
Field (JSON key) & Description\\
\midrule
Total resources (\texttt{nb\_res}) & The maximum number of compute nodes a job could request.\\
Job ID (\texttt{job\_id}) & Unique identifier of the job.\\
Requested resources (\texttt{res}) & Number of resources (e.g., nodes) requested by the job.\\
Submission time (\texttt{subtime}) & Job submission time.\\
User ID (\texttt{user\_id}) & Identifier of the user submitting the job.\\
Requested wall-time (\texttt{reqtime}) & Maximum runtime requested by the user.\\
Actual runtime (\texttt{runtime}) & Realized execution time of the job.\\
Profile (\texttt{profile}) & Identifier of the job's execution profile.\\
        \bottomrule
    \end{tabular}
\end{table}

Users should turn their real problem cases into these formats by encoding the target platform in a \texttt{platform.json} file and the job stream in a \texttt{workload.json} file. Additionally, SPARS includes a workload generator that can be configured to produce a JSON file of workloads with tunable characteristics. Users can control how frequently jobs arrive (arrival rate), the average execution time and its variability, the minimum and maximum number of nodes requested per job, and the number of jobs.

\subsubsection{Configurations}\label{sec:configs}

The simulator's behavior is controlled through configuration files, allowing experiments to be changed without modifying the source code. A runtime YAML file (\texttt{simulator\_config.yaml}) governs the main options---workload, platform, output path, scheduling algorithm, and timing policies.
A second configuration layer, defined in \texttt{Config.py}, controls the GYM interface. This includes registries for feature extractors, action translators, rewards, and learners. Together, these configuration layers make it easy to explore alternative policies and objectives in a reproducible manner. Samples of the configuration files are available within the GitHub repository.

\subsubsection{Running the simulation}\label{isect6}

Experiments are executed through a driver script (e.g., \texttt{runner.py}) that loads \texttt{simulator\_config.yaml}, constructs the \texttt{Simulator}, and then runs either a standard non-RL simulation or an RL-based simulation through a GYM-compatible environment. In the non-RL case, the simulator processes job arrivals, executions, completions, and node-state transitions while invoking the selected scheduling policy at each decision point. When RL is enabled, the simulator is wrapped in \texttt{HPCGymEnv}, allowing an agent to observe the system state, choose 
actions, and receive reward feedback defined by the user. In both modes, SPARS produces CSV outputs including job execution logs, node-state traces, and aggregate statistics such as total energy consumption, wasted energy, and waiting-time summaries.

\subsection{Software setup}\label{isect7}

To set up SPARS, users can first clone the source code from the GitHub repository using the command in \hyperref[lst:clone-repo]{Listing \ref{lst:clone-repo}}.

\begin{lstlisting}[
  language=sh,
  caption={Command to clone the SPARS repository.},
  label={lst:clone-repo}
]
git clone https://github.com/RakaSP/SPARS-Pub.git
\end{lstlisting}

SPARS supports both Linux and Windows. Users on either platform should create and activate a Python virtual environment, then install the required packages. Detailed setup instructions for each operating system are available in the project's GitHub repository.

\section{Illustrative examples}\label{isect8}

To demonstrate SPARS's capabilities, we present simulation experiments conducted on representative workloads and platform configurations. We use workload traces from the NASA Ames iPSC/860, CIEMAT Euler, and CEA Curie systems, all sourced from the Parallel Workload Archive \cite{1573950400806284800}. For each workload trace, the platform's number of compute nodes is set to the maximum requested job size in that trace. The detailed parameters used in this simulation are summarized in \hyperref[params]{Table~\ref{params}}. Note that we use a simplified static power model for illustration.

\begin{table}[ht]
    \centering
    \caption{Representative HPC platform and workload parameters used in the illustrative example.}
    \label{params}
    \begin{tabular}{l l}
        \toprule
        Parameter & Value \\
        \midrule
        Number of compute nodes & 64 (CIEMAT), 128 (NASA), 11200 (CEA) \\
        Compute speed & 1 \\
        Node active power & 190 W \\
        Transition-on power & 190 W \\
        Transition-on time & 30 minutes \\
        Transition-off power & 9 W \\
        Transition-off time & 45 minutes \\
        Sleeping power & 9 W \\
        Workload trace & \begin{tabular}[t]{@{}l@{}}
            CIEMAT Euler (last 1{,}000 jobs)
             \\NASA Ames iPSC/860 (last 10{,}839 jobs)
             \\
            CEA Curie (last 1{,}000 jobs)
        \end{tabular} \\
        \bottomrule
    \end{tabular}
\end{table}

SPARS also supports DVFS and richer multi-state power models, which can be configured via the platform JSON file. However, we do not evaluate DVFS in this paper because, to the best of our knowledge, there is no publicly available HPC workload trace that provides sufficient information for accurate simulation of DVFS. At minimum, such a trace would likely need to include (i) fine-grained per-node temperature measurements over time and (ii) platform-specific thermal-governor rules (e.g., temperature thresholds) that determine when DVFS should be triggered for each job.

In the experiments, we intentionally configure long transition delays (30 minutes for switch-on and 45 minutes for switch-off) to represent a more challenging and consequential scenario in which naive shutdown policies can severely increase queueing delay. Furthermore, this configuration aligns with prior work \cite{budiarjo2025improving,khasyah2022advantage}, allowing direct comparison of scheduler performance across studies.

Six schedulers are evaluated using SPARS, representing all combinations of two base policies: FCFS and EASY Backfilling, and three of our PSM variants developed in~\cite{workshop}:
\begin{itemize}
    \item \textbf{PSUS}: determines node availability based solely on reservation status, ignoring actual power states, and triggers switch-on upon node reservation.
    \item \textbf{PSAS (Auto On)}: similar to PSUS, but also considers actual node power states during allocation.
    \item \textbf{PSAS+IPM}: integrates scheduling and PSM as a unified policy, proactively waking sleeping nodes ahead of demand and selecting node combinations that minimize energy waste and transition overhead.
\end{itemize}
We refer to these six schedulers as: FCFS PSUS, EASY PSUS, FCFS PSAS (Auto On), EASY PSAS (Auto On), FCFS PSAS+IPM, and EASY PSAS+IPM.

\subsection{Scheduler analysis and performance metrics}\label{isect9}

One of the most direct ways to analyze scheduling performance in SPARS is through Gantt chart visualization. Gantt charts provide a temporal view of job execution across compute nodes, making it possible to inspect utilization efficiency, resource fragmentation, and the effects of scheduling and power-state decisions on performance and energy usage. They reveal when nodes are sleeping, idle, switching on and off, and executing jobs.

\begin{figure}[!ht]
  \centering
  \includegraphics[width=0.95\linewidth]{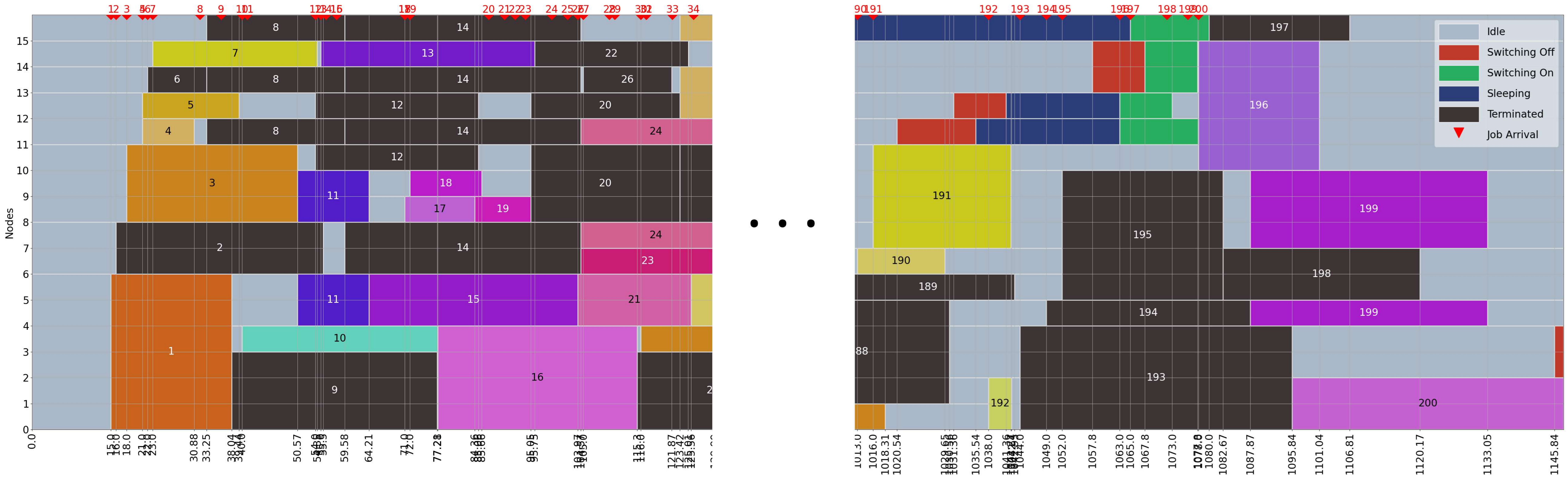}
  \caption{Example Gantt chart produced by SPARS for a workload of 200 jobs on 16 nodes, using the EASY Backfilling PSUS with a 50-second timeout policy and a terminate overrun policy.}
  \label{fig:gantt}
\end{figure}

\hyperref[fig:gantt]{Fig.~\ref{fig:gantt}} presents a Gantt chart produced by SPARS for a workload of randomly generated 200 jobs executed on 16 nodes under the EASY PSUS scheduler with a 50-second timeout shutdown policy and terminate overrun policy. In this setting, the timeout shutdown policy shuts down nodes after they have remained idle for 50 seconds, while the terminate overrun policy terminates jobs that exceed their requested wall-time. The horizontal axis of the figure represents simulation time, while the vertical axis corresponds to compute nodes. Colored blocks labeled with a job ID at their center represent job executions, while black blocks represent terminated jobs. Light blue blocks indicate idle periods, dark blue blocks indicate sleeping nodes, red blocks indicate switching-off transitions, and green blocks indicate switching-on transitions. This visualization enables researchers to intuitively examine scheduling dynamics, identify potential inefficiencies, and observe how node-state transitions affect job executions and idle periods.

\begin{figure}[!ht]
  \centering
   \includegraphics[width=0.9\linewidth]{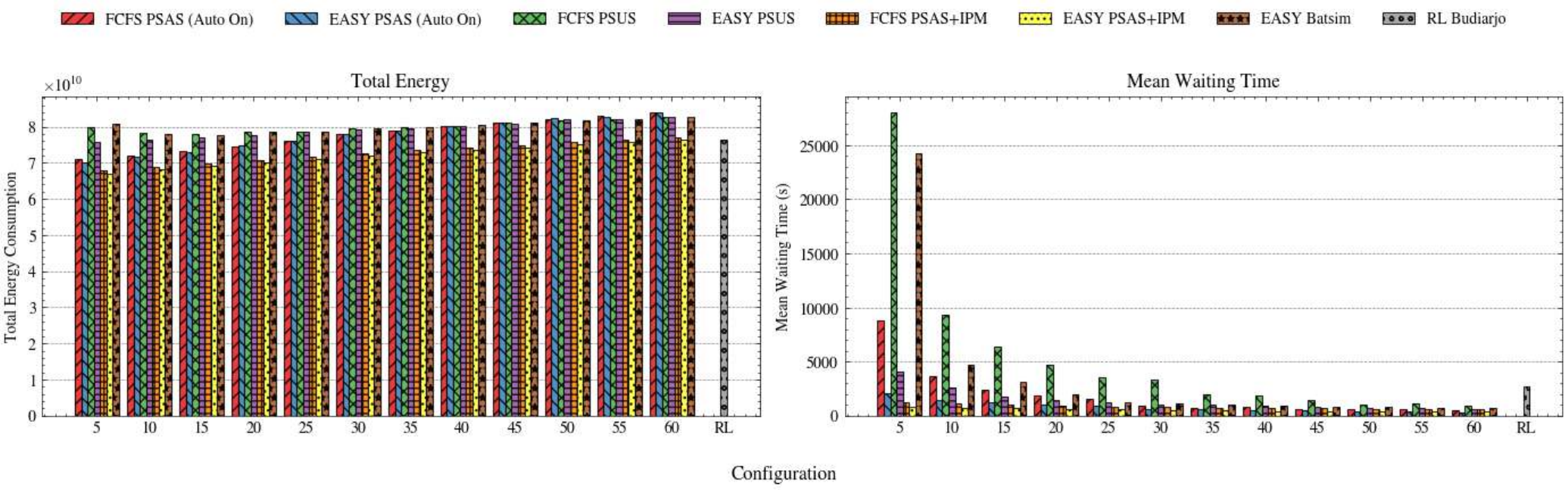}
   \caption{Comparison of cumulative energy consumption and average job waiting time across six schedulers on the NASA Ames iPSC/860 workload.}
   \label{fig:ecwt_comparison}
\end{figure}

\hyperref[fig:ecwt_comparison]{Fig.~\ref{fig:ecwt_comparison}} summarizes the cumulative energy consumption and average job waiting time for the six aforementioned schedulers implemented in SPARS, each evaluated with a timeout-based shutdown policy ranging from 5 to 60 minutes. For comparison, the figure also includes EASY Batsim, i.e., the EASY PSUS policy implemented in Batsim under the same timeout settings, and RL Budiarjo, an EASY Batsim variant with RL-based PSM proposed by Budiarjo et al.~in \cite{budiarjo2025improving}. All results are evaluated on the NASA Ames iPSC/860 workload.
%where EASY PSUS denotes SPARS's implementation of EASY Backfilling with the PSUS PSM policy and EASY Batsim denotes the same EASY PSUS policy implemented in Batsim.

To assess SPARS's reliability as a simulator, we compare the metrics obtained from EASY PSUS, implemented in SPARS, and EASY Batsim, implemented in Batsim, as shown in \hyperref[fig:ecwt_comparison]{Fig.~\ref{fig:ecwt_comparison}}. For Batsim, we use Batsim v4.1.0 with the \texttt{batsim-py} interface. We include this Batsim--SPARS comparison as a validation step: when running the same workload, platform configuration, and scheduling policy, SPARS should produce closely matching results, which increases confidence that SPARS is correct. The results show only a 1\% deviation in total energy consumption. This deviation may stem from the Batsim event-processing issues discussed earlier.

Having established that SPARS produces the correct simulation results, we now evaluate simulation performance. We run matched experiments in SPARS and Batsim on the CIEMAT Euler workload trace under the EASY Backfilling PSUS policy, sweeping timeout values from 5 to 60 minutes. All runs are executed in the same environment: a machine equipped with an Intel Core i5-11400H CPU (6 cores and 12 threads), an NVIDIA GeForce RTX 3050 GPU, 24 GB of RAM, and a 512 GB Samsung NVMe SSD. The experiments were conducted on Ubuntu 24.04.1 LTS (Noble) running under WSL2 on Windows 11. The remainder of the paper uses the same setup.

On the CIEMAT Euler workload trace under the EASY PSUS and EASY Batsim policies, SPARS delivers a substantial runtime improvement. \hyperref[tab:runtime-breakdown-easy]{Table~\ref{tab:runtime-breakdown-easy}} shows that, across all 12 timeout settings, SPARS requires only 1.86--2.8\% of \texttt{batsim-py}'s runtime, corresponding to a speedup of 35.5$\times$--53.5$\times$. On average, SPARS reduces runtime by 49.4~seconds per run and achieves a 44.54$\times$ speedup. These results indicate that SPARS is a much more efficient platform for evaluating scheduler policies. %More detailed performance-energy trade-offs and sensitivity analyses are reported in our companion study~\cite{workshop}.

\begin{table*}[!ht]
  \centering
  \caption{Runtime breakdown (counts and seconds) for Batsim-py (bat) and SPARS (spr) across timeout settings, evaluated on the CIEMAT Euler workload.}
  \label{tab:runtime-breakdown-easy}
  \resizebox{\textwidth}{!}{%
  \begin{tabular}{ll rr rr rr rr rr rr rr rr rr}
    \toprule
    \multirow{2}{*}{Timeout} & \multirow{2}{*}{Sim}
      & \multicolumn{2}{c}{Sim Advance}
      & \multicolumn{2}{c}{Scheduling}
      & \multicolumn{2}{c}{Resource}
      & \multicolumn{2}{c}{Job Lifecycle}
      & \multicolumn{2}{c}{Monitoring}
      & \multicolumn{2}{c}{Comm/IO}
      & \multicolumn{2}{c}{Timeout Pol.}
      & \multicolumn{2}{c}{Others}
      & \multicolumn{2}{c}{Total} \\
    \cmidrule(lr){3-4}\cmidrule(lr){5-6}\cmidrule(lr){7-8}\cmidrule(lr){9-10}\cmidrule(lr){11-12}\cmidrule(lr){13-14}\cmidrule(lr){15-16}\cmidrule(lr){17-18}\cmidrule(lr){19-20}
      & & Count & s & Count & s & Count & s & Count & s & Count & s & Count & s & Count & s & Count & s & Count & s \\
    \midrule

    300  & bat & 4,508,085 & 0.724 & 473,366 & 5.001 & 3,865,550 & 43.928 & 5,464,305 & 2.628 & 71,060 & 1.661 & 103,754 & 3.740 & 386,672 & 0.054 & 13,272,742 & 4.230 & 3,977,422 & 61.968 \\
    300  & spr &   462,972 & 0.063 & 460,476 & 0.432 &   603,584 & 0.026 &   945,205 & 0.002 & 734,615 & 0.274 &     N/A & 0.000 &  63,150 & 0.043 &  8,165,723 & 0.318 & 2,632,730 & 1.158 \\
    \midrule

    1200 & bat & 4,463,179 & 0.714 & 539,247 & 4.145 & 4,100,280 & 38.012 & 5,139,618 & 2.198 & 68,115 & 1.704 & 119,677 & 3.163 & 409,862 & 0.050 & 13,915,748 & 3.701 & 4,183,465 & 53.688 \\
    1200 & spr &   262,426 & 0.115 & 357,924 & 0.513 &   388,303 & 0.039 &   516,787 & 0.004 & 513,107 & 0.384 &     N/A & 0.000 &  43,815 & 0.061 &  6,092,333 & 0.398 & 1,885,964 & 1.514 \\
    \midrule

    2100 & bat & 5,285,914 & 0.596 & 645,031 & 3.409 & 4,586,539 & 33.178 & 5,944,326 & 1.850 & 78,857 & 1.458 & 122,133 & 3.084 & 442,467 & 0.046 & 16,164,808 & 3.110 & 4,692,964 & 46.733 \\
    2100 & spr &   468,421 & 0.063 & 455,165 & 0.403 &   623,544 & 0.024 &   969,007 & 0.002 & 691,749 & 0.281 &     N/A & 0.000 &  53,564 & 0.049 &  7,510,737 & 0.319 & 2,473,568 & 1.141 \\
    \midrule

    3000 & bat & 5,122,590 & 0.610 & 629,424 & 3.483 & 4,567,038 & 33.045 & 5,735,592 & 1.915 & 76,356 & 1.481 & 121,285 & 3.086 & 374,210 & 0.054 & 14,667,898 & 3.403 & 4,623,064 & 47.076 \\
    3000 & spr &   483,759 & 0.057 & 479,238 & 0.380 &   647,693 & 0.022 & 1,064,361 & 0.002 & 705,313 & 0.258 &     N/A & 0.000 &  53,517 & 0.046 &  7,600,305 & 0.302 & 2,533,564 & 1.067 \\
    \midrule

  \end{tabular}%
  }
\end{table*}

The step-level breakdown in \hyperref[tab:runtime-breakdown-easy]{Table~\ref{tab:runtime-breakdown-easy}} shows that this speedup is driven mainly by lower resource overhead and the elimination of Comm/IO costs. Averaged across runs, resource operations decrease from 35.71~seconds in \texttt{batsim-py} to 0.03~seconds in SPARS, while Comm/IO overhead is effectively removed. Overall, SPARS gains most of its advantage from lightweight resource and power-state handling, together with the removal of communication-related overheads.

To evaluate larger-scale performance, we also simulate the last 1{,}000 jobs of the CEA Curie workload trace on a platform with 11{,}200 nodes. SPARS completed the simulation in 312~seconds, compared to 17{,}992~seconds for Batsim-py, corresponding to roughly 57$\times$ speedup. This shows that SPARS maintains its runtime advantage at larger scales and is especially useful for RL workflows that require many repeated simulations.

\begin{figure}[!ht]
  \centering
   \includegraphics[width=0.9\linewidth]{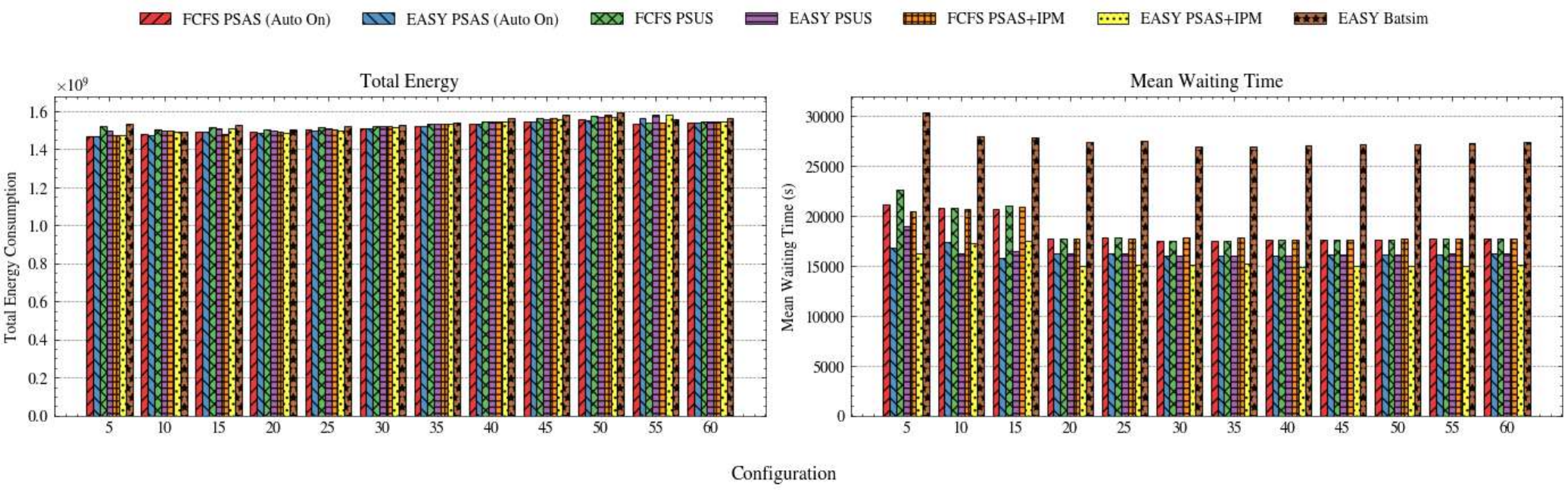}
   \caption{Comparison of cumulative energy consumption and average job waiting time across six schedulers on the CIEMAT Euler workload.}
   \label{fig:ecwt_comparison_ciemat_euler}
\end{figure}

\hyperref[fig:ecwt_comparison_ciemat_euler]{Fig.~\ref{fig:ecwt_comparison_ciemat_euler}} further validates SPARS on the more recent CIEMAT Euler workload, using the last 1{,}000 jobs from December 2017. The total energy consumption between EASY Batsim and EASY PSUS differs by only 0.75\% on average across all timeout settings, consistent with the NASA trace results. This confirms that SPARS accurately reproduces scheduling and energy behavior across different workloads.

\subsection{Reproducibility and extensibility}\label{isect10}

The configuration files, source code, and plotting scripts used to produce the reported figures are publicly available in the project's GitHub repository---ensuring that all experiments can be replicated and extended. Furthermore, users may adapt the workload and platform JSON files to model alternative systems.\enlargethispage{-4pt}

\section{Impact}\label{isect11}

SPARS lowers the barrier to evaluating HPC batch scheduling with power management and RL-based policies by providing a lightweight simulator with built-in support for PSM and a native RL interface. This makes it straightforward to assess how power policies and RL-based approaches affect system-level outcomes such as energy consumption and job waiting time under controlled workload and platform assumptions, enabling researchers and practitioners to refine scheduling strategies before deployment on real HPC infrastructure.

SPARS also opens important research opportunities. As shown by \cite{workshop}, RL-based power managers can underperform when they are paired with schedulers whose autonomous power decisions override the agent's actions, degrading the reward signal and learned policy. This highlights the need for RL agents that are explicitly designed to coordinate with power-state-aware schedulers. More broadly, SPARS's modular and extensible design makes it well-suited for studying energy--performance trade-offs, developing adaptive scheduling and PSM policies, testing new scheduling heuristics, and exploring scalability under larger workloads and future HPC systems.

Beyond research, SPARS has practical value for HPC operators and institutions that rely on production clusters. By simulating realistic workload conditions before deployment, users can evaluate scheduling strategies without costly trial-and-error on real systems. Its extensible design also allows future integration of richer hardware models, such as node-level power curves, cooling overheads, and DVFS, making it useful for studying energy-aware strategies that reduce operational costs, emissions, and hardware stress. SPARS has already supported the evaluation of PSAS+IPM \cite{workshop}, a co-designed scheduler and power manager that outperformed conventional timeout-based and RL-driven approaches in simulation, while reported production deployments of selective node power-down policies suggest that such evaluations can translate into substantial energy savings \cite{suarez2025energy}.\enlargethispage{-15pt}

\section{Conclusion}\label{isect12}

SPARS is a lightweight, reproducible simulator for studying HPC batch scheduling together with node PSM, with optional RL-based control via a Gym-compatible interface. By processing scheduling decisions and power transitions within a single event-driven engine, SPARS supports systematic evaluation of energy--performance trade-offs under \nobreak{}configurable workloads, platforms, and policy variants, while remaining easy to extend with new heuristics, observation/reward designs, and learning algorithms. We validate SPARS against Batsim on matched workloads and configurations, showing close agreement in aggregate energy metrics and demonstrating that SPARS can execute the same experiments substantially faster through reduced resource-bookkeeping overhead and elimination of communication-related costs. This speed advantage is particularly beneficial for RL workflows that require many repeated simulations. Future work includes incorporating calibrated platform-specific power--performance data to evaluate DVFS-aware policies.

\section*{Acknowledgements}

This work is partly supported by {Universitas Gadjah Mada's Program Asistensi Riset 2025 No.~4588/UN1.P4/Dit-Lit/PT.01.03/2025 and JSPS KAKENHI} Grant Number JP24K02945.

\section*{CRediT authorship contribution statement}
\textbf{M. A. Amrizal}: Writing -- review \& editing, Writing -- original draft, Visualization, Validation, Supervision, Software, Methodology, Investigation, Formal analysis, Conceptualization.
\textbf{R. S. Prasasta}: Writing -- review \& editing, Writing -- original draft, Visualization, Software, Methodology, Investigation, Formal analysis.
\textbf{S. Y. Pradata}: Writing -- review \& editing, Writing -- original draft, Methodology, Investigation, Formal analysis.
\textbf{K. G. Santiyuda}: Writing -- review \& editing, Validation, Methodology, Formal analysis. 
\textbf{R. Pulungan}: Writing -- review \& editing, Validation, Supervision, Resources, Methodology, Funding acquisition, Formal analysis, Conceptualization.
\textbf{H. Takizawa}: Writing -- review \& editing, Validation, Methodology, Formal analysis.

\section*{Declaration of competing interest}
All authors declare that they have no known competing financial
interests or personal relationships that could have appeared to
influence the work reported in this paper.

\section*{Data and implementation}
We have created a repository to store the source code and data in
\url{https://github.com/RakaSP/SPARS-Pub}.

\bibliographystyle{elsarticle-num} 
\bibliography{mybibliography}

@InProceedings{khasyah2022advantage,
author="Khasyah, Fitra Rahmani
and Santiyuda, Kadek Gemilang
and Kaunang, Gabriel
and Makhrus, Faizal
and Amrizal, Muhammad Alfian
and Takizawa, Hiroyuki",
editor="Takizawa, Hiroyuki
and Shen, Hong
and Hanawa, Toshihiro
and Hyuk Park, Jong
and Tian, Hui
and Egawa, Ryusuke",
title="An Advantage Actor-Critic Deep Reinforcement Learning Method for Power Management in {HPC} Systems",
booktitle="Parallel and Distributed Computing, Applications and Technologies",
year="2023",
publisher="Springer Nature Switzerland",
address="Cham",
pages="94--107",
isbn="978-3-031-29927-8",
doi = {10.1007/978-3-031-29927-8_8}
}

@book{buyya2013mastering,
author = {Buyya, Rajkumar and Vecchiola, Christian and Selvi, S. Thamarai},
title = {Mastering Cloud Computing: Foundations and Applications Programming},
year = {2013},
isbn = {9780124095397},
publisher = {Morgan Kaufmann Publishers Inc.},
address = {San Francisco, CA, USA},
edition = {1st},
doi = {10.1016/C2012-0-06719-1}
}

@inproceedings{Ohmura,
author = {Ohmura, Tatsuyoshi and Shimomura, Yoichi and Egawa, Ryusuke and Takizawa, Hiroyuki},
title = {Toward Building a Digital Twin of Job Scheduling and Power Management on an {HPC} System},
year = {2022},
isbn = {978-3-031-22697-7},
publisher = {Springer-Verlag},
address = {Berlin, Heidelberg},
opturl = {https://doi.org/10.1007/978-3-031-22698-4_3},
doi = {10.1007/978-3-031-22698-4_3},
booktitle = {Job Scheduling Strategies for Parallel Processing: 25th International Workshop, JSSPP 2022, Virtual Event, June 3, 2022, Revised Selected Papers},
pages = {47–67},
numpages = {21}
}

@inproceedings{budiarjo2025improving,
author = {Budiarjo, Thomas and Pradata, Santana Yuda and Santiyuda, Kadek Gemilang and Amrizal, Muhammad Alfian and Pulungan, Reza and Takizawa, Hiroyuki},
title = {Improving the Efficiency of a Deep Reinforcement Learning-Based Power Management System for {HPC} Clusters Using Curriculum Learning},
year = {2025},
isbn = {9798400712500},
publisher = {Association for Computing Machinery},
address = {New York, NY, USA},
opturl = {https://doi.org/10.1145/3718350.3718359},
doi = {10.1145/3718350.3718359},
booktitle = {Proceedings of the 2025 Supercomputing Asia Conference},
pages = {1--13},
numpages = {13},
location = {
},
series = {SCA '25}
}

@InProceedings{dutot2015batsim,
author="Dutot, Pierre-Fran{\c{c}}ois
and Mercier, Michael
and Poquet, Millian
and Richard, Olivier",
editor="Desai, Narayan
and Cirne, Walfredo",
title="Batsim: A Realistic Language-Independent Resources and Jobs Management Systems Simulator",
booktitle="Job Scheduling Strategies for Parallel Processing",
year="2017",
publisher="Springer International Publishing",
address="Cham",
pages="178--197",
isbn="978-3-319-61756-5",
doi={10.1007/978-3-319-61756-5_10}
}

@INPROCEEDINGS{carastan2019one,
  author={Carastan-Santos, Danilo and De Camargo, Raphael Y. and Trystram, Denis and Zrigui, Salah},
  booktitle={2019 19th IEEE/ACM International Symposium on Cluster, Cloud and Grid Computing (CCGRID)}, 
  title={One Can Only Gain by Replacing {EASY} Backfilling: A Simple Scheduling Policies Case Study}, 
  year={2019},
  volume={},
  number={},
  pages={1-10},
  doi={10.1109/CCGRID.2019.00010}}

@INPROCEEDINGS{feitelson1998utilization,
  author={Feitelson, D.G. and Weil, A.M.},
  booktitle={Proceedings of the First Merged International Parallel Processing Symposium and Symposium on Parallel and Distributed Processing}, 
  title={Utilization and predictability in scheduling the {IBM SP2} with backfilling}, 
  year={1998},
  volume={},
  number={},
  pages={542-546},
  doi={10.1109/IPPS.1998.669970}}

@misc{batsim-issue67,
  title   = {Simultaneous events sent in different {Batsim} messages},
  howpublished = {\url{https://github.com/oar-team/batsim/issues/67}},
  note    = {{GitHub} issue \#67 (opened Dec 15, 2021; closed)},
  year    = {2021},
  author  = {Maël Madon and others}
}

@inproceedings{workshop,
author = {Prasasta, Raka Satya and Pradata, Santana Yuda and Santiyuda, Kadek Gemilang and Amrizal, Muhammad Alfian and Pulungan, Reza and Takizawa, Hiroyuki},
title = {Co-Design of a Power State-Aware Scheduler and an Intelligent Power Manager for Energy-Efficient {HPC} Systems},
year = {2026},
isbn = {9798400723285},
publisher = {Association for Computing Machinery},
address = {New York, NY, USA},
opturl = {https://doi.org/10.1145/3784828.3785163},
doi = {10.1145/3784828.3785163},
booktitle = {Proceedings of the Supercomputing Asia and International Conference on High Performance Computing in Asia Pacific Region Workshops},
pages = {13–21},
numpages = {9},
location = {
},
series = {SCA/HPCAsiaWS '26}
}

@techreport{nrdc,
  author      = "Josh Whitney and Pierre Delforge",
  title       = "Data Center Efficiency Assessment",
  institution = "Natural Resources Defense Council (NRDC)",
  type        = {Issue Paper},
  number      = {IP:14-08-a},
  year        = {2014},
  url = "https://www.nrdc.org/sites/default/files/data-center-efficiency-assessment-IP.pdf"
}

@book{suttonbarto,
author = {Sutton, Richard S. and Barto, Andrew G.},
title = {Reinforcement Learning: An Introduction},
year = {2018},
isbn = {0262039249},
publisher = {A Bradford Book},
address = {Cambridge, MA, USA}
}

@article{fan2021dras,
title = {{DRAS-CQSim}: A reinforcement learning based framework for {HPC} cluster scheduling},
journal = {Software Impacts},
volume = {8},
pages = {100077},
year = {2021},
issn = {2665-9638},
doi = {10.1016/j.simpa.2021.100077},
opturl = {https://www.sciencedirect.com/science/article/pii/S2665963821000257},
author = {Yuping Fan and Zhiling Lan}
}

@INPROCEEDINGS{zhang2020rlscheduler,
  author={Zhang, Di and Dai, Dong and He, Youbiao and Bao, Forrest Sheng and Xie, Bing},
  booktitle={SC20: International Conference for High Performance Computing, Networking, Storage and Analysis}, 
  title={{RLScheduler}: An Automated {HPC} Batch Job Scheduler Using Reinforcement Learning}, 
  year={2020},
  volume={},
  number={},
  pages={1-15},
  doi={10.1109/SC41405.2020.00035}}

@Article{kocot2023energy,
AUTHOR = {Kocot, Bartłomiej and Czarnul, Paweł and Proficz, Jerzy},
TITLE = {Energy-Aware Scheduling for High-Performance Computing Systems: A Survey},
JOURNAL = {Energies},
VOLUME = {16},
YEAR = {2023},
NUMBER = {2},
ARTICLE-NUMBER = {890},
optURL = {https://www.mdpi.com/1996-1073/16/2/890},
ISSN = {1996-1073},
DOI = {10.3390/en16020890}
}

@ARTICLE{suarez2025energy,
AUTHOR={Suarez, Estela  and Bockelmann, Hendryk  and Eicker, Norbert  and Eitzinger, Jan  and El Sayed, Salem  and Fieseler, Thomas  and Frank, Martin  and Frech, Peter  and Giesselmann, Pay  and Hackenberg, Daniel  and Hager, Georg  and Herten, Andreas  and Ilsche, Thomas  and Koller, Bastian  and Laure, Erwin  and Manzano, Cristina  and Oeste, Sebastian  and Ott, Michael  and Reuter, Klaus  and Schneider, Ralf  and Thust, Kay  and von St. Vieth, Benedikt },
TITLE={Energy-aware operation of {HPC} systems in {Germany}},
JOURNAL={Frontiers in High Performance Computing},
          
VOLUME={Volume 3 - 2025},
  
YEAR={2025},
  
optURL={https://www.frontiersin.org/journals/high-performance-computing/articles/10.3389/fhpcp.2025.1520207},
  
DOI={10.3389/fhpcp.2025.1520207},
  
ISSN={2813-7337}}

@article{galleguillos2020accasim,
author = {Galleguillos, Cristian and Kiziltan, Zeynep and Netti, Alessio and Soto, Ricardo},
title = {{AccaSim}: A customizable workload management simulator for job dispatching research in {HPC} systems},
year = {2020},
issue_date = {Mar 2020},
publisher = {Kluwer Academic Publishers},
address = {USA},
volume = {23},
number = {1},
issn = {1386-7857},
opturl = {https://doi.org/10.1007/s10586-019-02905-5},
doi = {10.1007/s10586-019-02905-5},
journal = {Cluster Computing},
month = mar,
pages = {107–122},
numpages = {16}
}

@inproceedings{ozden2022elastisim,
author = {\"{O}zden, Taylan and Beringer, Tim and Mazaheri, Arya and Fard, Hamid Mohammadi and Wolf, Felix},
title = {{ElastiSim}: A Batch-System Simulator for Malleable Workloads},
year = {2023},
isbn = {9781450397339},
publisher = {Association for Computing Machinery},
address = {New York, NY, USA},
opturl = {https://doi.org/10.1145/3545008.3545046},
doi = {10.1145/3545008.3545046},
booktitle = {Proceedings of the 51st International Conference on Parallel Processing},
articleno = {40},
numpages = {11},
location = {Bordeaux, France},
series = {ICPP '22}
}

@article{kahil2025reinforcement,
title = {Reinforcement learning for data center energy efficiency optimization: A systematic literature review and research roadmap},
journal = {Applied Energy},
volume = {389},
pages = {125734},
year = {2025},
issn = {0306-2619},
doi = {10.1016/j.apenergy.2025.125734},
opturl = {https://www.sciencedirect.com/science/article/pii/S0306261925004647},
author = {Hussain Kahil and Shiva Sharma and Petri Välisuo and Mohammed Elmusrati}
}

@misc{1573950400806284800,
  author       = {Feitelson, Dror G.},
  title        = {{Parallel Workloads Archive}},
  howpublished = {\url{https://www.cs.huji.ac.il/labs/parallel/workload}},
  note         = {Accessed: December 7, 2025},
  year = {2025}
}

@InProceedings{yoo2003slurm,
author="Yoo, Andy B.
and Jette, Morris A.
and Grondona, Mark",
editor="Feitelson, Dror
and Rudolph, Larry
and Schwiegelshohn, Uwe",
title="{SLURM}: Simple Linux Utility for Resource Management",
booktitle="Job Scheduling Strategies for Parallel Processing",
year="2003",
publisher="Springer Berlin Heidelberg",
address="Berlin, Heidelberg",
pages="44--60",
isbn="978-3-540-39727-4"
}

@misc{1606.01540,
      title={{OpenAI Gym}}, 
      author={Greg Brockman and Vicki Cheung and Ludwig Pettersson and Jonas Schneider and John Schulman and Jie Tang and Wojciech Zaremba},
      year={2016},
      eprint={1606.01540},
      archivePrefix={arXiv},
      primaryClass={cs.LG},
      url={https://arxiv.org/abs/1606.01540}, 
}

@INPROCEEDINGS{8906698,
  author={Agung, Mulya and Amrizal, Muhammad Alfian and Egawa, Ryusuke and Takizawa, Hiroyuki},
  booktitle={2019 IEEE 13th International Symposium on Embedded Multicore/Many-core Systems-on-Chip (MCSoC)}, 
  title={An Automatic {MPI} Process Mapping Method Considering Locality and Memory Congestion on {NUMA} Systems}, 
  year={2019},
  volume={},
  number={},
  pages={17-24},
  doi={10.1109/MCSoC.2019.00010}
}

@ARTICLE{9239288,
  author={Lin, Jianpeng and Cui, Delong and Peng, Zhiping and Li, Qirui and He, Jieguang},
  journal={IEEE Access}, 
  title={A Two-Stage Framework for the Multi-User Multi-Data Center Job Scheduling and Resource Allocation}, 
  year={2020},
  volume={8},
  number={},
  pages={197863-197874},
  doi={10.1109/ACCESS.2020.3033557}
}

@ARTICLE{9698981,
  author={Zeng, Jing and Ding, Ding and Kang, Kaixuan and Xie, HuaMao and Yin, Qian},
  journal={IEEE Transactions on Parallel and Distributed Systems}, 
  title={Adaptive {DRL}-Based Virtual Machine Consolidation in Energy-Efficient Cloud Data Center}, 
  year={2022},
  volume={33},
  number={11},
  pages={2991-3002},
  doi={10.1109/TPDS.2022.3147851}
}

@ARTICLE{8772127,
  author={Li, Yuanlong and Wen, Yonggang and Tao, Dacheng and Guan, Kyle},
  journal={IEEE Transactions on Cybernetics}, 
  title={Transforming Cooling Optimization for Green Data Center via Deep Reinforcement Learning}, 
  year={2020},
  volume={50},
  number={5},
  pages={2002-2013},
  doi={10.1109/TCYB.2019.2927410}
}

\end{document}